\documentclass[twocolumn,showpacs,preprintnumbers]{revtex4}%
\usepackage{amssymb}
\usepackage{amsmath}
\usepackage{graphicx}
\usepackage{dcolumn}
\usepackage{bm}
\usepackage{epstopdf}
\usepackage{amsfonts}%
\begin{document}
\title{Self-excited oscillation and synchronization of an on-fiber optomechanical cavity}
\author{Eyal Buks}
\affiliation{Andrew and Erna Viterbi Department of Electrical Engineering, Technion, Haifa
32000 Israel}
\author{Ivar Martin}
\affiliation{Materials Science Division, Argonne National Laboratory, Argonne, Illinois
60439, USA}
\date{\today }

\begin{abstract}
We study a fully on-fiber optomechanical cavity and characterize its
performance as a sensor. The cavity is formed by patterning a suspended
metallic mirror near the tip of an optical fiber and by introducing a static
reflector inside the fiber. Optically induced self-excited oscillation (SEO) is
observed above a threshold value of the injected laser power. The SEO phase
can be synchronized by periodically modulating the optical power that is
injected into the cavity. Noise properties of the system in the region of
synchronization are investigated. Moreover, the spectrum is measured near different values of the modulation frequency, at which phase locking occurs. A universal behavior is revealed in the transition between the regions of phase locked and free running SEO.

\end{abstract}
\pacs{03.65.Yz, 05.40.-a, 42.50.Pq}
\maketitle





\section{Introduction}

Resonant detection is a widely employed technique in a variety of
applications. A detector belonging to this class typically consists of a
resonator, which is characterized by an angular resonance frequency $\omega_{0}$
and characteristic damping rate $\gamma_{0}$. Detection is achieved by coupling
a physical parameter of interest, denoted here as $q$, to the resonator
in such a way that $\omega_{0}$ becomes  $q$ dependent. The sensitivity
of the detection scheme that is employed for monitoring the parameter of
interest $q$ can be characterized by the minimum detectable change in $q$,
denoted as $\delta_{q}$. For small changes, $\delta_{q}$ is related to the
 minimum detectable {\em relative} change in the frequency $\sigma_{\mathrm{\omega}%
}=\delta_{\omega_{0}}/\omega_{0}$ by the relation $\delta_{q}=\left\vert
\partial\omega_{0}/\partial q\right\vert ^{-1}\omega_{0}\sigma_{\mathrm{\omega}}$. The
dimensionless parameter $\sigma_{\mathrm{\omega}}$, in turn, typically depends on
the noises affecting the resonator and on the averaging time $t_{\mathrm{a}}$ of the measurement.%

\begin{figure}
[ptb]
\begin{center}
\includegraphics[
height=3.6019in,
width=3.4541in
]%
{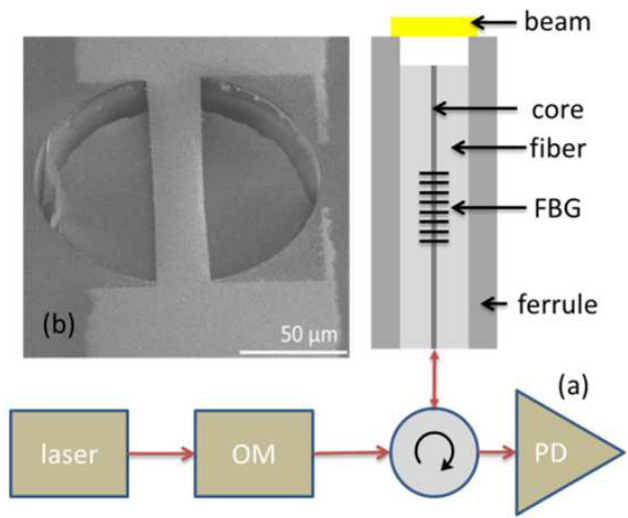}%
\caption{Experimental setup. (a) A schematic drawing of the sample and the
experimental setup. An on-fiber optomechanical cavity is excited by a tunable
laser with external optical modulator (OM). The mechanical resonator has
quality factor $\omega_{0}/2\gamma_{0}=3800$ and the cavity has finesse
$\beta_{\mathrm{F}}=2.1$. The reflected light intensity is measured using a
photodetector (PD), which is connected to both a spectrum analyzer and an
oscilloscope (not shown in the sketch). (b) Electron micrograph of the
suspended micromechanical mirror, whose mass is $m=1.1\times10^{-12}%
\operatorname{kg}$.}%
\label{FigSetup}%
\end{center}
\end{figure}

A commonly employed detection scheme is based on externally driving the
resonator with a monochromatic force at a frequency close to the resonance
frequency and monitoring the response using homodyne detection. For this case
the normalized minimum detectable change in the frequency is found to be given
by $\sigma_{\mathrm{\omega}}=\sigma_{\mathrm{{\omega}0}}$, where \cite{Cleland_2758}%
\begin{equation}
\sigma_{\mathrm{{\omega}0}}=\left(  \frac{2\gamma_{0} k_{\mathrm{B}}T_{\mathrm{eff}}%
}{U_{0}\omega_{0}^{2}t_{\mathrm{a}}}\right)  ^{1/2}\ , \label{sigma_y ext}%
\end{equation}
$k_{\mathrm{B}}$ is the Boltzmann's constant, $T_{\mathrm{eff}}$ is the noise
effective temperature and $U_{0}$ is the energy stored in the externally
driven resonator in steady state. Note that Eq. (\ref{sigma_y ext}) is derived
by assuming that the response of the resonator is linear and by assuming the
classical limit, i.e. $k_{\mathrm{B}}T_{\mathrm{eff}}\gg\hbar\omega_{0}$, where
$\hbar$ is Planks's constant. The generalization of Eq. (\ref{sigma_y ext})
for the case of nonlinear response is discussed in Ref. \cite{Buks_026217}.
Note that other contributions to $\sigma_{\mathrm{\omega}}$ (e.g. instrumental
noise) are not analyzed in this paper.

In the just mentioned scheme the resonator is driven into a so-called state
of forced oscillation (FO) by applying a fixed periodic external force.
Alternatively, in some cases back-reaction effects can be exploited for
generating a so-called state of self-excited oscillation (SEO)
\cite{Rugar1989, Arcizet2006a, Forstner2012,Weig2013}. In this paper we study
SEO in an on-fiber optomechanical cavity device (see Fig. \ref{FigSetup}),
which is formed between a fiber Bragg grating (FBG) mirror, serving as a
static reflector, and a vibrating mirror fabricated on a mechanical resonator next to  the tip of
the single mode optical fiber. 
The movable mirror can be driven into the SEO state by injecting a monochromatic laser light into
the fiber. The SEO occurs via bolometric feedback effects. Further, by periodically modulating the
laser intensity it is possible to lock the  phase of SEO to a rational fraction of the modulation frequency.

Driving an on-fiber optomechanical cavity into the state of SEO is easier in comparison with the case of FO, which requires the fabrication of additional elements near the tip of the fiber that need to be electrically wired. In contrast, SEO can be induced by optical excitation only. In this paper, we study the performance of the device as a sensor in the regime of SEO, both  
with and without synchronization, and compare the results to the case of
FO [see Eq. (\ref{sigma_y ext})]. We find that synchronization gives rise to a suppression in phase noise, which in turn allows a sensitivity enhancement.

Optomechanical cavities \cite{Braginsky_653,
Hane_179,Gigan_67,Metzger_1002,Kippenberg_1172,Favero_104101,Marquardt2009}
are widely employed for various sensing \cite{Rugar1989, Arcizet2006a,
Forstner2012,Weig2013} and photonics
\cite{Lyshevski&Lyshevski_03,Stokes_et_al_90, Hossein_Zadeh_276,Wu_et_al_06,
MattEichenfield2007,Bahl2011,FlowersJacobs_221109} applications. Such systems
may allow experimental study of the crossover between classical to quantum
realms \cite{Poot_273}.
 The effect of radiation pressure typically governs the
optomechanical coupling (i.e. the coupling between the electromagnetic cavity
and the mechanical resonator that serves as a movable mirror) when the finesse
of the optical cavity is sufficiently high
\cite{Kippenberg_et_al_05,Rokhsari2005,
Arcizet2006,Gigan_67,Cooling_Kleckner06, Kippenberg_1172}, whereas, bolometric
effects can contribute to the optomechanical coupling when optical absorption
by the vibrating mirror is significant \cite{Metzger_1002,
Jourdan_et_al_08,Marino&Marin2011PRE, Metzger_133903, Restrepo_860,
Liberato_et_al_10,Marquardt_103901, Paternostro_et_al_06,Yuvaraj_430}.
Generally, bolometric effects are dominant in systems comprising of relatively
large mirrors, in which the thermal relaxation rate is comparable to the
mechanical resonance frequency \cite{Aubin_1018, Marquardt_103901,
Paternostro_et_al_06, Liberato_et_al_10_PRA}. These systems
\cite{Metzger_133903, Metzger_1002, Aubin_1018,
Jourdan_et_al_08,Zaitsev_046605,Zaitsev_1589} exhibit many intriguing
phenomena including mode cooling
\cite{Hane_179,Kim_1454225,Aubin_1018,Carmon_223902,Marquardt_103901,Corbitt_021802,Carmon_123901,Metzger_133903}
and synchronization \cite{Shlomi_032910,Wang_023007}.

\section{Experimental setup}

The optomechanical cavity, which is schematically shown in Fig. \ref{FigSetup}%
(a), was fabricated on the flat polished tip of a single mode fused silica
optical fiber with outer diameter of $126%
\operatorname{\mu m}%
$ (Corning SMF-28 operating at wavelength band around $1550%
\operatorname{nm}%
$) held in a zirconia ferrule \cite{Yuvaraj_210403}. A $10%
\operatorname{nm}%
$ thick chromium layer and a $200%
\operatorname{nm}%
$ gold layer were successively deposited by thermal evaporation. The bilayer
was directly patterned by a focused ion beam to the desired mirror shape ($20%
\operatorname{\mu m}%
$ wide doubly clamped beam). Finally, the mirror was released by etching
approximately $12%
\operatorname{\mu m}%
$ of the underlying silica in 7\% HF acid ($90%
\operatorname{min}%
$ etch time at room temperature). The suspended mirror remained supported by
the zirconia ferrule, which is resistant to HF.

The static mirror of the optomechanical cavity was provided by a FBG mirror
\cite{Zaitsev_046605} (made using a standard phase mask technique
\cite{Anderson_566}, with grating period of $0.527%
\operatorname{\mu m}%
$ and length $\approx8%
\operatorname{mm}%
$) having reflectivity band of $0.4%
\operatorname{nm}%
$ full width at half maximum centered at $\lambda_{0}=1545%
\operatorname{nm}%
$. The length of the optical cavity was $l\approx10%
\operatorname{mm}%
$, providing a free spectral range of $\Delta\lambda=\lambda_{0}%
^{2}/2n_{\mathrm{eff}}l\approx80$ pm (where $n_{\mathrm{eff}}$ $=1.468$ is the
effective refraction index for SMF-28).

Monochromatic light was injected into the fiber bearing the cavity on its tip
from a laser source with an adjustable output wavelength $\lambda$ and power
level $P_{\mathrm{L}}$. The laser was connected through an optical circulator,
that allowed the measurement of the reflected light intensity $P_{\mathrm{R}}$
by a fast responding photodetector [see Fig. \ref{FigSetup}(a)]. The detected
signal was analyzed by an oscilloscope and a spectrum analyzer. The
experiments were performed in vacuum (at residual pressure below $0.01%
\operatorname{Pa}%
$) at a base temperature of $77%
\operatorname{K}%
$.

The optically-induced SEO in our device is attributed to the bolometric
optomechanical coupling between the optical mode and the mechanical resonator
\cite{Zaitsev_046605,Zaitsev_1589}. While the phase of SEO randomly diffuses
in time when the laser power that is injected into the cavity is kept
constant, phase locking
\cite{Anishchenko_117,Pandey_3,Paciorek_1723,Adler_351,Jensen_1637,DosSantos_1147}
may occur when the laser power is periodically modulated in time. Such locking
results in entrainment \cite{Hamerly_1504_04410}, i.e. synchronization
\cite{huygens1986pendulum,Rosenblum_401} between the SEO and the external
modulation \cite{Georg_043603}.

\section{Evolution equation}

In the limit of small displacement the dynamics of the system can be
approximately described using a first order evolution equation given by
\cite{Zaitsev_1589,Dykman_1646}%
\begin{equation}
\dot{A}+\left(  \Gamma_{\mathrm{eff}}+i\Omega_{\mathrm{eff}}\right)
A=\xi\left(  t\right)  +\vartheta\left(  t\right)  \;. \label{A dot D}%
\end{equation}
The complex amplitude $A$ is related to the mechanical displacement $x\left(
t\right)  $ by $x\left(  t\right)  =x_{0}+2\operatorname{Re}A$, where $x_{0}$
is an optically-induced static displacement. Overdot denotes a derivative with
respect to time $t$. To lowest nonvanishing order in $\left\vert A\right\vert
^{2}$ the damping rate $\Gamma_{\mathrm{eff}}$ and the angular resonance
frequency $\Omega_{\mathrm{eff}}$ are given by $\Gamma_{\mathrm{eff}}%
=\Gamma_{0}+\Gamma_{2}\left\vert A\right\vert ^{2}$ and $\Omega_{\mathrm{eff}%
}=\Omega_{0}+\Omega_{2}\left\vert A\right\vert ^{2}$. The term $\xi\left(
t\right)  $ represents the effective force that is generated due to the laser
power modulation. The fluctuating term $\vartheta\left(  t\right)
=\vartheta_{x}\left(  t\right)  +i\vartheta_{y}\left(  t\right)  $, where both
$\vartheta_{x}$ and $\vartheta_{y}$ are real, represents white noise
\cite{Risken_Fokker-Planck,Fong_023825} and the following is assumed to hold:
$\left\langle \vartheta_{x}\left(  t\right)  \vartheta_{x}\left(  t^{\prime
}\right)  \right\rangle =\left\langle \vartheta_{y}\left(  t\right)
\vartheta_{y}\left(  t^{\prime}\right)  \right\rangle =2\Theta\delta\left(
t-t^{\prime}\right)  $ and $\left\langle \vartheta_{x}\left(  t\right)
\vartheta_{y}\left(  t^{\prime}\right)  \right\rangle =0$, where
$\Theta=\gamma_{0}k_{\mathrm{B}}T_{\mathrm{eff}}/4m\omega_{0}^{2}$,
$\gamma_{0}$ and $\omega_{0}$ are, respectively, the intrinsic damping rate
and angular frequency of the resonator, $m$ is its mass, $k_{\mathrm{B}}$ is
the Boltzmann's constant and $T_{\mathrm{eff}}$ is the effective noise
temperature. Expressions for the coefficients $x_{0}$, $\Gamma_{0}$,
$\Gamma_{2}$, $\Omega_{0}$ and $\Omega_{2}$, which all depend on the
properties of the optical cavity and on the laser wavelength and power, are
given in appendix \ref{App_EE}.

\section{Resonance detection with SEO}

In the absence of laser modulation, i.e. when $\xi\left(  t\right)  =0$, the
equation of motion (\ref{A dot D}) describes a van der Pol oscillator
\cite{Pandey_3}. Consider the case where $\Gamma_{2}>0$, for which a
supercritical Hopf bifurcation occurs when the linear damping coefficient
$\Gamma_{0}$ vanishes. Above threshold, i.e. when $\Gamma_{0}$ becomes
negative, the amplitude $A_{r}$ of SEO is given by $r_{0}=\sqrt{-\Gamma
_{0}/\Gamma_{2}}$ and the angular frequency $\Omega_{\mathrm{H}}$ of SEO by
$\Omega_{\mathrm{H}}=\Omega_{\mathrm{eff}}\left(  r_{0}\right)  $.

The method of resonance detection can be implemented in the regime of SEO. The
normalized minimum detectable change in the frequency $\sigma_{\mathrm{\omega}}$ in
this regime has been evaluated in \cite{Baskin_563}, and it was found to be
given by%
\begin{equation}
\sigma_{\mathrm{\omega}}=\sigma_{\mathrm{{\omega}}0}\left(  1+\frac{\zeta_{0}^{2}%
}{4\left\vert \Gamma_{0}\right\vert \Gamma_{2}}\right)  ^{1/2}\;,
\label{sigma_A SEO}%
\end{equation}
where $\zeta_{0}=\mathrm{d}\Omega_{\mathrm{eff}}/\mathrm{d}\left\vert
A\right\vert $. The above result (\ref{sigma_A SEO}) indicates that for the
same value of the stored energy $U_{0}$ the smallest detectable change
$\delta_{q}$ in the measured parameter $q$\ is $\left(  1+\zeta_{0}%
^{2}/4\left\vert \Gamma_{0}\right\vert \Gamma_{2}\right)  ^{1/2}$ larger for
the current case of SEO in comparison with the case of FO [see Eq.
(\ref{sigma_y ext})]. The reduced sensitivity is attributed to the dependence
of $\Omega_{\mathrm{eff}}$ on the amplitude $\left\vert A\right\vert $ of
oscillation, which gives rise to elevated phase noise. Recall that in the
regime of SEO, in which no periodic driving is applied, the phase of
oscillation is not externally dictated, and consequently it becomes more
susceptible to noise. For the device under study in this work the degradation
factor is given by $\left(  1+\zeta_{0}^{2}/4\left\vert \Gamma_{0}\right\vert
\Gamma_{2}\right)  ^{1/2}\simeq10$.

\section{Synchronization}

In the regime of SEO phase noise can be suppressed by modulating the injected
laser power at an angular frequency $\omega_{\mathrm{d}}$ close to the angular
frequency $\Omega_{\mathrm{eff}}$. When fluctuations are dominated by phase
noise Eq. (\ref{A dot D}) can be simplified by disregarding fluctuations in
the amplitude $\left\vert A\right\vert $ (i.e. by assuming that $\left\vert
A\right\vert =r_{0}$). In this approach one finds that the relative phase
$\gamma$ between mechanical oscillation and the applied modulation evolves in
time according to [see Eq. (\ref{CBJJ V2})]%
\begin{equation}
\frac{\mathrm{d}\gamma}{\mathrm{d}\tau}+\sin\gamma=i_{\mathrm{b}%
}+i_{\mathrm{n}}\;, \label{CBJJ V1}%
\end{equation}
where $i_{\mathrm{b}}=\left(  \omega_{\mathrm{d}}-\Omega_{\mathrm{eff}%
}\right)  /\omega_{\mathrm{a}}$ is a normalized detuning, $\tau=\omega
_{\mathrm{a}}t$ is a dimensionless time variable and $\omega_{\mathrm{a}}$ is
the modulation amplitude. The term $i_{\mathrm{n}}$ represents white noise
having a vanishing expectation value $\left\langle i_{\mathrm{n}}\right\rangle
=0$ and autocorrelation function $\left\langle i_{\mathrm{n}}\left(
\tau\right)  i_{\mathrm{n}}\left(  \tau^{\prime}\right)  \right\rangle
=\left(  2\Theta/\omega_{\mathrm{a}}r_{0}^{2}\right)  \delta\left(  \tau
-\tau^{\prime}\right)  $ [see Eq. (\ref{<i_n(tau)*i_n(tau')>})].

In the region of synchronization, in which $\left\vert i_{\mathrm{b}%
}\right\vert \leq1$, Eq. (\ref{CBJJ V1}) has a stationary solution given by
$\gamma=\sin^{-1}i_{\mathrm{b}}$, whereas $\gamma$ becomes time dependent when
$\left\vert i_{\mathrm{b}}\right\vert >1$. In that region and when noise is
disregarded the time evolution of $\gamma\left(  \tau\right)  $, which is
given by Eq. (\ref{gamma(tau)}), is a periodic function of $\tau$ with a
period given by $T_{\mathrm{J}}=2\pi/\sqrt{i_{\mathrm{b}}^{2}-1}$ [see Eq.
(\ref{CBJJ T})]. The Fourier series expansion of $\mathrm{d}\gamma
/\mathrm{d}\tau$ is given by Eqs. (\ref{V FE}) and (\ref{g_k}). The periodic
time evolution of the relative phase $\gamma$ gives rise to sidebands in the
spectrum at the angular frequencies $\omega_{\mathrm{d}}+n\omega_{\mathrm{s}}%
$, where $n$ is an integer and the sideband spacing $\omega_{\mathrm{s}}$ is
given by $\omega_{\mathrm{s}}=2\pi\omega_{\mathrm{a}}/T_{\mathrm{J}}%
=\sqrt{\left(  \omega_{\mathrm{d}}-\Omega_{\mathrm{eff}}\right)  ^{2}%
-\omega_{\mathrm{a}}^{2}}$.

The equation of motion (\ref{CBJJ V1}) indicates that the dynamics of the
relative phase $\gamma$ is governed by a potential $U_{\mathrm{b}}\left(
\gamma\right)  =-\cos\gamma-i_{\mathrm{b}}\gamma$ having the shape of a tilted
washboard. The shape of the barriers separating local minima points of
$U_{\mathrm{b}}\left(  \gamma\right)  $ (when $\left\vert i_{\mathrm{b}%
}\right\vert <1$) can be controlled by adjusting the frequency detuning
$\omega_{\mathrm{d}}-\Omega_{\mathrm{eff}}$ and modulation amplitude
$\omega_{\mathrm{a}}$. The highly nonlinear response of the system near to
onset point of synchronization can be exploited for some sensing applications.
Near synchronization threshold, i.e. when $\left\vert i_{\mathrm{b}%
}\right\vert \lesssim1$, noise may give rise to phase slip events (i.e.
transitions between neighboring potential wells). The average rate of
noise-induced events can be estimated using the Kramers formula
\cite{Kramers_284,Hanggi_251}. Note that an equation of motion similar to
(\ref{CBJJ V1}) governs the dynamics of a current-biased Josephson junction in
the so-called overdamped regime \cite{Tinkham1975}. The effect of the noise
term $i_{\mathrm{n}}$ on voltage fluctuations across a Josephson junction in
the quantum regime has been investigated in
\cite{Kogan_ElectronicNoise,Levinson_184504}.

\section{SA measurements}

\begin{figure}
[ptb]
\begin{center}
\includegraphics[
height=3.6871in,
width=3.4546in
]%
{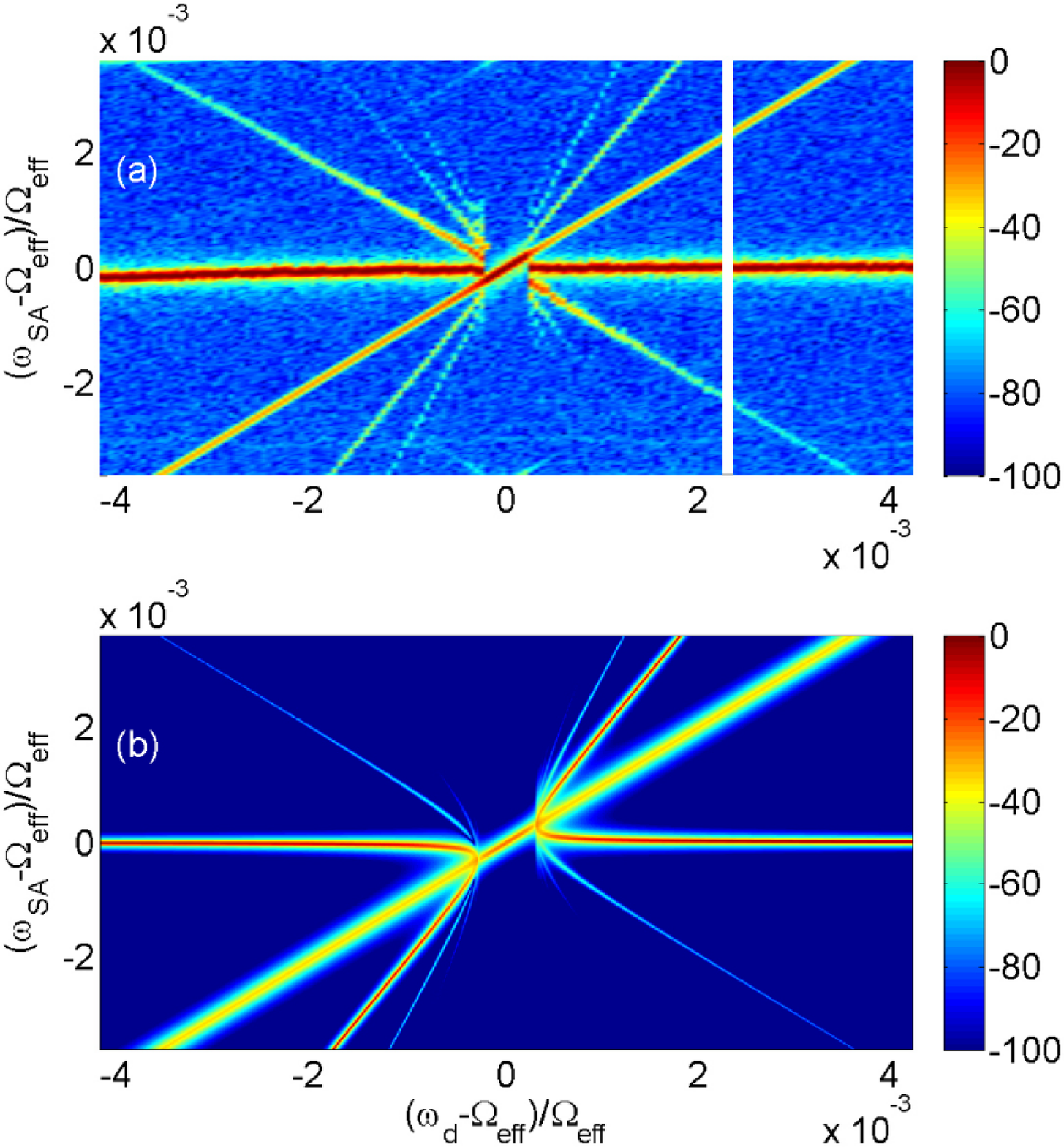}%
\caption{Power spectrum (in dB units) vs. detuning. (a) The measured power spectrum as a
function of $\left(  \omega_{\mathrm{d}}-\Omega_{\mathrm{eff}}\right)
/\Omega_{\mathrm{eff}}$, where $\Omega_{\mathrm{eff}}/2\pi
=236.4\operatorname{kHz}$. In this measurement the average laser power is set
to $P_{\mathrm{L}}=12\operatorname{mW}$, the wavelength to $\lambda
=1545.498\operatorname{nm}$ and the dimensionless laser modulation amplitude
to $0.025$. (b) Theoretical calculation of the power spectrum based on the
solution of Eq. (\ref{CBJJ V1}). In both plots synchronization occurs when
$\left\vert i_{\mathrm{b}}\right\vert <1$ and sidebands are observed when
$\left\vert i_{\mathrm{b}}\right\vert >1$.}%
\label{FigSA}%
\end{center}
\end{figure}

The measured power spectrum of the photodetector signal is plotted in Fig.
\ref{FigSA}(a) as a function of $\left(  \omega_{\mathrm{d}}-\Omega
_{\mathrm{eff}}\right)  /\Omega_{\mathrm{eff}}=i_{\mathrm{b}}\omega
_{\mathrm{a}}/\Omega_{\mathrm{eff}}$. In the absence of power laser modulation
the frequency of SEO is given by $\Omega_{\mathrm{eff}}/2\pi=236.4%
\operatorname{kHz}%
$. As can be seen from Fig. \ref{FigSA}(a), in the region $\left\vert \left(
\omega_{\mathrm{d}}-\Omega_{\mathrm{eff}}\right)  /\Omega_{\mathrm{eff}%
}\right\vert \leq2.1\times10^{-4}$ synchronization occurs, and thus
$\omega_{\mathrm{a}}/2\pi=49.6%
\operatorname{Hz}%
$ for this measurement. The above-discussed side bands are clearly visible in
the region $\left\vert \left(  \omega_{\mathrm{d}}-\Omega_{\mathrm{eff}%
}\right)  /\Omega_{\mathrm{eff}}\right\vert >\omega_{\mathrm{a}}%
/\Omega_{\mathrm{eff}}$, i.e. when $\left\vert i_{\mathrm{b}}\right\vert >1$.
The theoretically calculated power spectrum is presented in Fig.
\ref{FigSA}(b) for comparison. The amplitudes of the sidebands are determined
using the Fourier series expansion [see Eqs. (\ref{V FE}) and (\ref{g_k})].
Good agreement between data and theory is found.%

\begin{figure}
[ptb]
\begin{center}
\includegraphics[
height=2.8052in,
width=3.4537in
]%
{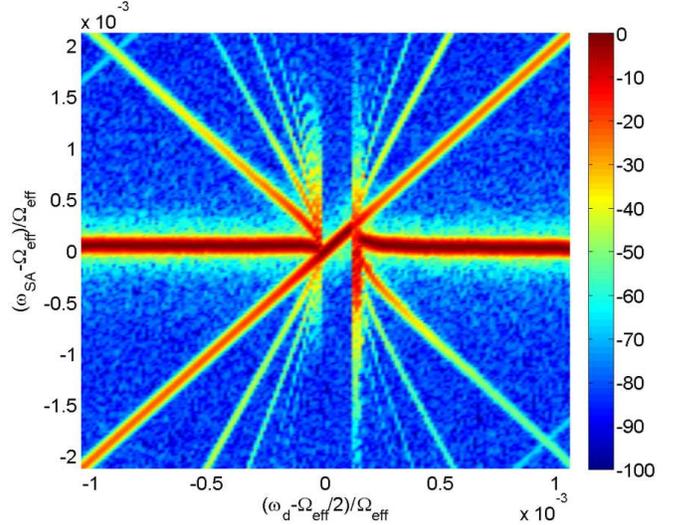}%
\caption{Measured power spectrum (in dB units) vs. detuning for the case where
$\omega_{\mathrm{d}}/\Omega_{\mathrm{eff}}\simeq1/2$. Other experimental
parameters are listed in the caption of Fig. \ref{FigSA}.}%
\label{FigSA_F}%
\end{center}
\end{figure}

For the data presented in Fig. \ref{FigSA}(a) synchronization occurs when
$\omega_{\mathrm{d}}/\Omega_{\mathrm{eff}}\simeq1$. Similar synchronization is
observed when the ratio $\omega_{\mathrm{d}}/\Omega_{\mathrm{eff}}$ is tuned
close to other rational values \cite{Shlomi_032910,Wang_023007}. As an
example, the measured power spectrum for the case where $\omega_{\mathrm{d}%
}/\Omega_{\mathrm{eff}}\simeq1/2$ is shown in Fig. \ref{FigSA_F}. As can be
seen from the comparison with Fig. \ref{FigSA}(a), the results are
qualitatively similar. Not shown here, but very similar behaviors were also experimentally observed
at other ratios including $\omega_{\mathrm{d}}/\Omega_{\mathrm{eff}}%
\simeq1/3$, $2/3$, $1/4$, $3/4$, $1/5$, $2/5$, $3/5$ and $4/5$. 
This highlights universal aspects of the phase-locking phenomenon when the drive frequency approximately matches a rational multiple of the natural self-oscillation frequency.

Phase locking near rational values of the ratio $\omega_{\mathrm{d}}%
/\Omega_{\mathrm{eff}}$ is qualitatively discussed in appendix \ref{App_SRL}.
It is found that just outside the locking regime the frequency of oscillation
undergoes a transition from a value dictated by the externally applied
modulation to the value corresponding to free running SEO. Moreover, this
transition is expected to obey a  ``square root law" for both integer and fractional
values of the ratio $\omega_{\mathrm{d}}/\Omega_{\mathrm{eff}}$. That is, the unlocking occurs continuously, with the deviation between the oscillator frequency from the (fraction of) drive frequency depending on the control parameter $\delta$ (drive amplitude, or detuning) as $\propto \sqrt{|\delta - \delta_c|}$, where $\delta_c$ is the critical -- ``unlocking" -- value of $\delta$.
This
universality in the behavior just outside locking regime is demonstrated by
the similarity of the spectrum measured near different locking regions (see
Figs. \ref{FigSA} and \ref{FigSA_F}).

\section{Resonance detection in the region of synchronization}

The effect of noise, which is assumed to be weak, can be taken into account by linearizing Eq.
(\ref{CBJJ V1}), which becomes%
\begin{equation}
\frac{\mathrm{d}\gamma_{\mathrm{n}}}{\mathrm{d}\tau}+\gamma_{\mathrm{n}}%
\cos\gamma=i_{\mathrm{n}}\;, \label{d gamma_n/d tau}%
\end{equation}
where $\gamma$ represents a solution in the noiseless case and $\gamma
_{\mathrm{n}}$ is a fluctuating term added to $\gamma$ due to noise. For the
case $\left\vert i_{\mathrm{b}}\right\vert <1$ one finds from the stationary
solution of Eq. (\ref{CBJJ V1}) that $\cos\gamma=\sqrt{1-i_{\mathrm{b}}^{2}}$. The power spectrum $S_{\gamma
}\left(  w\right)  $ of $\gamma_{\mathrm{n}}$ as a function of the
dimensionless angular frequency $w$ can be found using Eq.
(\ref{d gamma_n/d tau}) and the autocorrelation function $\left\langle
i_{\mathrm{n}}\left(  \tau\right)  i_{\mathrm{n}}\left(  \tau^{\prime}\right)
\right\rangle $%
\begin{equation}
S_{\gamma}\left(  w\right)  =\frac{\Theta}{\pi\omega_{\mathrm{a}}r_{0}^{2}%
}\frac{1}{1-i_{\mathrm{b}}^{2}+w^{2}}\;.
\end{equation}
The correlation function $C_{\gamma}\left(  \tau^{\prime}\right)
=\left\langle \gamma_{\mathrm{n}}\left(  \tau+\tau^{\prime}\right)
\gamma_{\mathrm{n}}\left(  \tau\right)  \right\rangle $ can be determined
using the Wiener-Khinchine theorem%
\begin{equation}
C_{\gamma}\left(  \tau^{\prime}\right)  =\int_{-\infty}^{\infty}%
\mathrm{d}w\;e^{iw\tau^{\prime}}S_{\gamma}\left(  w\right)  =\frac{\Theta
e^{-\sqrt{1-i_{\mathrm{b}}^{2}}\left\vert \tau^{\prime}\right\vert }}%
{\omega_{\mathrm{a}}r_{0}^{2}\sqrt{1-i_{\mathrm{b}}^{2}}}\;,
\end{equation}
and thus the following holds $\left\langle \exp\left(  i\left(  \gamma
_{\mathrm{n}}\left(  \tau+\tau^{\prime}\right)  -\gamma_{\mathrm{n}}\left(
\tau\right)  \right)  \right)  \right\rangle =1-\gamma_{\mathrm{g}}\left\vert
\tau^{\prime}\right\vert +O\left(  \left\vert \tau^{\prime}\right\vert
^{2}\right)  $, where $\gamma_{\mathrm{g}}$, which is given by $\gamma
_{\mathrm{g}}=\Theta/\omega_{\mathrm{a}}r_{0}^{2}$, is the jitter rate. Note
that $\gamma_{\mathrm{g}}$ is independent on $i_{\mathrm{b}}$. In terms of the
the energy $U_{0}$ stored in the mechanical resonator the dimensionfull jitter
rate is given by $\omega_{\mathrm{a}}\gamma_{\mathrm{g}}=\left(
k_{\mathrm{B}}T_{\mathrm{eff}}/U_{0}\right)  \gamma_{0}$. Note that the same
jitter rate is obtained for the case of FO (provided that $U_{0}$ is unchanged).

In the region of synchronization, i.e. when $\left\vert i_{\mathrm{b}%
}\right\vert <1$, the resonance frequency $\Omega_{\mathrm{eff}}$ can be
determined by monitoring the relative phase $\gamma$. Consider an estimate of
the averaged value of $\gamma$ based on a sampling of the signal
$\gamma+\gamma_{\mathrm{n}}\left(  \tau\right)  $ over a sampling time of
duration $\tau_{\mathrm{a}}$. The standard deviation $\delta_{\gamma}$ in this
estimate is given by%
\begin{equation}
\delta_{\gamma}=\left(  \frac{2\pi S_{\gamma}\left(  0\right)  }%
{\tau_{\mathrm{a}}}\right)  ^{1/2}=\left(  \frac{2\Theta}{\omega_{\mathrm{a}%
}\tau_{\mathrm{a}}r_{0}^{2}\left(  1-i_{\mathrm{b}}^{2}\right)  }\right)
^{1/2}\;.
\end{equation}
The corresponding deviation $\delta_{\Omega}$ in the estimated value of
$\Omega_{\mathrm{eff}}$ is given by $\delta_{\Omega}=\mathcal{R}\delta
_{\gamma}$, where the responsivity $\mathcal{R}$ is given by $\mathcal{R}%
=\left(  \left\vert \mathrm{d}\gamma/\mathrm{d}\Omega_{\mathrm{eff}%
}\right\vert \right)  ^{-1}=\omega_{\mathrm{a}}\sqrt{1-i_{\mathrm{b}}^{2}}$ ,
and thus the normalized minimum detectable change in the frequency is given by
$\delta_{\Omega}/\Omega_{\mathrm{H}}=\sigma_{\mathrm{{\omega}0}}$, where
$\sigma_{\mathrm{{\omega}0}}$ is given by Eq. (\ref{sigma_y ext}). Thus for the same
given stored energy $U_{0}$ in the mechanical resonator the same sensitivity
is expected in the regions of synchronization and FO. In other words, the
above-discussed degradation in sensitivity occurring in the region of SEO [see
Eq. (\ref{sigma_A SEO})] can be fully eliminated by synchronization-induced
suppression of phase noise. Similarly to the case of FO, in the regime of synchronization the phase of mechanical oscillation is externally dictated, and consequently noise is suppressed.

\section{Summary}

In summary, the sensitivity of a detector based on a mechanical resonator is studied in the SEO regime. It is found that phase noise can be suppressed by externally applying a modulation. Moreover, the transition from the regions of synchronization and free running SEO is explored, and a universal behavior referred to as the square-root law is revealed.

\section{Acknowledgements}

On-fiber optomechanical cavities have been fabricated by former members of the Technion group K. Shlomi, D. Yuvaraj
and I. Baskin. Argonne National Laboratory's contribution is based upon work supported by Laboratory Directed Research and Development (LDRD) funding from Argonne National Laboratory, provided by the Director, Office of Science, of the U.S. Department of Energy under Contract No. DE-AC02-06CH11357.

\appendix

\section{Evolution equation}

\label{App_EE}

The theoretical model \cite{Zaitsev_1589,Zaitsev_046605} that is used to
derive the evolution equation (\ref{A dot D}) is briefly described below. The
micromechanical mirror in the optical cavity is treated as a mechanical
resonator with a single degree of freedom $x$ having mass $m$ and linear
damping rate $\gamma_{0}$ (when it is decoupled from the optical cavity). It
is assumed that the angular resonance frequency of the mechanical resonator
depends on the temperature $T$ of the suspended mirror. For small deviation of
$T$ from the base temperature $T_{0}$ (i.e. the temperature of the supporting
substrate) it is taken to be given by $\omega_{0}-\beta T_{\mathrm{R}}$, where
$T_{\mathrm{R}}=T-T_{0}$ and where $\beta$ is a constant. Furthermore, to
model the effect of thermal deformation \cite{Metzger_133903} it is assumed
that a temperature dependent force given by $m\theta T_{\mathrm{R}}$, where
$\theta$ is a constant, acts on the mechanical resonator \cite{Yuvaraj_430}.
When noise is disregarded, the equation of motion governing the dynamics of
the mechanical resonator is taken to be given by%
\begin{equation}
\frac{\mathrm{d}^{2}x}{\mathrm{d}t^{2}}+2\gamma_{0}\frac{\mathrm{d}%
x}{\mathrm{d}t}+\left(  \omega_{0}-\beta T_{\mathrm{R}}\right)  ^{2}x=\theta
T_{\mathrm{R}}\;. \label{x eom}%
\end{equation}

The intra-cavity optical power incident on the suspended mirror is denoted by
$P_{\mathrm{L}}I\left(  x\right)  $, where $P_{\mathrm{L}}$ is the injected
laser power, and the function $I\left(  x\right)  $ depends on the mechanical
displacement $x$ [see Eq. (\ref{I(x) D}) below]. The time evolution of the
relative temperature $T_{\mathrm{R}}$ is governed by the thermal balance
equation%
\begin{equation}
\frac{\mathrm{d}T_{\mathrm{R}}}{\mathrm{d}t}=Q-\kappa T_{\mathrm{R}}\;,
\label{T_R eom}%
\end{equation}
where $Q=\eta P_{\mathrm{L}}I\left(  x\right)  $ is proportional to the
heating power, $\eta$ is the heating coefficient due to optical absorption and
$\kappa$ is the thermal decay rate.

The function $I\left(  x\right)  $ depends on the properties of the optical
cavity that is formed between the suspended mechanical mirror and the on-fiber
static reflector. The finesse of the optical cavity is limited by loss
mechanisms that give rise to optical energy leaking out of the cavity. The
main escape routes are through the on-fiber static reflector, through
absorption by the metallic mirror, and through radiation. The corresponding
transmission probabilities are respectively denoted by $\mathcal{T}%
_{\mathrm{B}}$, $\mathcal{T}_{\mathrm{A}}$ and $\mathcal{T}_{\mathrm{R}}$. In
terms of these parameters, the function $I\left(  x\right)  $ is given by
\cite{Zaitsev_046605}%
\begin{equation}
I\left(  x\right)  =\frac{\beta_{\mathrm{F}}\left(  1-\frac{\beta_{-}^{2}%
}{\beta_{+}^{2}}\right)  \beta_{+}^{2}}{1-\cos\frac{4\pi x_{\mathrm{D}}%
}{\lambda}+\beta_{+}^{2}}\;, \label{I(x) D}%
\end{equation}
where $x_{\mathrm{D}}=x-x_{\mathrm{R}}$ is the displacement of the mirror
relative to a point $x_{\mathrm{R}}$, at which the energy stored in the
optical cavity in steady state obtains a local maximum, $\beta_{\pm}%
^{2}=\left(  \mathcal{T}_{\mathrm{B}}\pm\mathcal{T}_{\mathrm{A}}\pm
\mathcal{T}_{\mathrm{R}}\right)  ^{2}/8$ and where $\beta_{\mathrm{F}}$ is the
cavity finesse. The reflection probability $R_{\mathrm{C}}=P_{\mathrm{R}%
}/P_{\mathrm{L}}$ is given in steady state by \cite{Yurke_5054,Zaitsev_046605}
$R_{\mathrm{C}}=1-I\left(  x\right)  /\beta_{\mathrm{F}}$. For sufficiently
small $x$, the expansion $I\left(  x\right)  =I_{0}+I_{0}^{\prime}x+\left(
1/2\right)  I_{0}^{\prime\prime}x^{2}+O\left(  x^{3}\right)  $ can be
employed, where a prime denotes differentiation with respect to the
displacement $x$.

Consider the case where the laser power $P_{\mathrm{L}}$ is modulated in time
according to $P_{\mathrm{L}}\left(  t\right)  =P_{0}+P_{1}\left(  t\right)  $,
where $P_{0}$ is a constant and $P_{1}\left(  t\right)  $ is assumed to have a
vanishing average. When both $P_{1}$ and $I-I_{0}$ are sufficiently small, the
problem can be significantly simplified by employing the approximation
$Q\simeq\eta P_{0}I+\eta P_{1}I_{0}$. The displacement $x\left(  t\right)  $
is expressed in terms of the complex amplitude $A$ as $x\left(  t\right)
=x_{0}+2\operatorname{Re}A$, where $x_{0}$, which is given by $x_{0}%
=\eta\theta P_{0}I_{0}/\kappa\omega_{0}^{2}$, is the averaged
optically-induced static displacement. For a small displacement, the evolution
equation for the complex amplitude $A$ is found to be given by Eq.
(\ref{A dot D})\ \cite{Zaitsev_1589,Shlomi_032910}, where both the effective
resonance frequency $\Omega_{\mathrm{eff}}$ and the effective damping rate
$\Gamma_{\mathrm{eff}}$ are real even functions of $\left\vert A\right\vert $.
To second order in $\left\vert A\right\vert $ they are given by $\Gamma
_{\mathrm{eff}}=\Gamma_{0}+\Gamma_{2}\left\vert A\right\vert ^{2}$ and
$\Omega_{\mathrm{eff}}=\Omega_{0}+\Omega_{2}\left\vert A\right\vert ^{2}$,
where $\Gamma_{0}=\gamma_{0}+\eta\theta P_{0}I_{0}^{\prime}/2\omega_{0}^{2}$,
$\Gamma_{2}=\gamma_{2}+\eta\beta P_{0}I_{0}^{\prime\prime}/4\omega_{0}$,
$\gamma_{2}$ is the intrinsic mechanical nonlinear quadratic damping rate
\cite{Zaitsev_859}, $\Omega_{0}=\omega_{0}-\eta\beta P_{0}I_{0}/\kappa$ and
$\Omega_{2}=-\eta\beta P_{0}I_{0}^{\prime\prime}/\kappa$. Note that the above
expressions for $\Gamma_{\mathrm{eff}}$ and $\Omega_{\mathrm{eff}}$ are
obtained by making the following assumptions: $\kappa^{2}/\omega_{0}%
^{3}\lambda\ll\beta/\theta\ll1/2\omega_{0}x_{0}$ and $\kappa\ll\omega_{0}$,
both typically hold experimentally \cite{Zaitsev_046605}. The term $\xi\left(
t\right)  $ in Eq. (\ref{A dot D}) represents the thermal force that is
generated due to the laser power modulation. For the case of a monochromatic
modulation at angular frequency $\omega_{\mathrm{d}}\simeq\Omega
_{\mathrm{eff}}$ it is given by $\xi\left(  t\right)  =\eta P_{1}I_{0}%
\theta\Omega_{0}^{-2}e^{-i\omega_{\mathrm{d}}t}$.

\section{Monochromatic modulation with $\omega_{\mathrm{d}}\simeq
\Omega_{\mathrm{eff}}$}

\label{App_MM}

In cylindrical coordinates the complex amplitude of mechanical oscillation $A$ is expressed as $A=A_{r}e^{iA_{\theta}}$, where
$A_{r}=\sqrt{A_{x}^{2}+A_{y}^{2}}$ is positive, $A_{\theta}$ is real
\cite{Hempstead_350}, and Eq. (\ref{A dot D}) is written as%
\begin{equation}
\dot{A}_{r}+A_{r}\Gamma_{\mathrm{eff}}=\xi_{r}+\vartheta_{r}\;,
\label{A_r dot}%
\end{equation}
and%
\begin{equation}
\dot{A}_{\theta}+\Omega_{\mathrm{eff}}\left(  A_{r}\right)  =\frac{\xi
_{\theta}+\vartheta_{\theta}}{A_{r}}\;, \label{A_theta dot}%
\end{equation}
where%
\begin{align}
\xi_{r}  &  =\frac{A_{x}\xi_{x}+A_{y}\xi_{y}}{A_{r}}\;,\label{xi_r}\\
\vartheta_{r}  &  =\frac{A_{x}\vartheta_{x}+A_{y}\vartheta_{y}}{A_{r}}\;,
\end{align}%
\begin{align}
\xi_{\theta}  &  =\frac{A_{x}\xi_{y}-A_{y}\xi_{x}}{A_{r}}\;,\label{xi_theta}\\
\vartheta_{\theta}  &  =\frac{A_{x}\vartheta_{y}-A_{y}\vartheta_{x}}{A_{r}}\;,
\end{align}
$A=A_{x}+iA_{y}$, $\xi=\xi_{x}+i\xi_{y}$, where $A_{x}$, $A_{y}$, $\xi_{x}$
and $\xi_{y}$ are all real, and the noise terms satisfy the following relations%
\begin{align}
\left\langle \vartheta_{r}\left(  t\right)  \vartheta_{r}\left(  t^{\prime
}\right)  \right\rangle  &  =2\Theta\delta\left(  t-t^{\prime}\right)
\;,\label{<xi_r xi_r>}\\
\left\langle \vartheta_{\theta}\left(  t\right)  \vartheta_{\theta}\left(
t^{\prime}\right)  \right\rangle  &  =2\Theta\delta\left(  t-t^{\prime
}\right)  \;,\label{<xi_theta xi_theta>}\\
\left\langle \vartheta_{r}\left(  t\right)  \vartheta_{\theta}\left(
t^{\prime}\right)  \right\rangle  &  =0\;. \label{<xi_r xi_theta>}%
\end{align}

Consider the case of a monochromatic modulation at angular frequency
$\omega_{\mathrm{d}}$, which is assumed to be close to the angular frequency
$\Omega_{\mathrm{eff}}$. When fluctuations in the amplitude $A_{r}$ of SEO are
disregarded, i.e. when it is assumed that $A_{r}=r_{0}$, Eq.
(\ref{A_theta dot}) becomes [see Eq. (\ref{xi_theta})]%
\begin{equation}
\dot{A}_{\theta}+\Omega_{\mathrm{eff}}=-\omega_{\mathrm{a}}\sin\left(
A_{\theta}+\omega_{\mathrm{d}}t\right)  +\frac{\vartheta_{\theta}}{r_{0}}\;,
\label{A_theta dot, Ar=r0}%
\end{equation}
where $\Omega_{\mathrm{eff}}\ $is treated as a constant and where
$\omega_{\mathrm{a}}$, which is given by $\omega_{\mathrm{a}}=\eta P_{1}%
I_{0}\theta/\Omega_{0}^{2}r_{0}$, represents the modulation amplitude. In
terms of the relative phase $\gamma=A_{\theta}+\omega_{\mathrm{d}}t$ and the
dimensionless time $\tau=\omega_{\mathrm{a}}t$ Eq. (\ref{A_theta dot, Ar=r0})
can be rewritten as%
\begin{equation}
\frac{\mathrm{d}\gamma}{\mathrm{d}\tau}+\sin\gamma=i_{\mathrm{b}%
}+i_{\mathrm{n}}\;, \label{CBJJ V2}%
\end{equation}
where%
\begin{equation}
i_{\mathrm{b}}=\frac{\omega_{\mathrm{d}}-\Omega_{\mathrm{eff}}}{\omega
_{\mathrm{a}}}\;, \label{i_b}%
\end{equation}
and where $i_{\mathrm{n}}=\vartheta_{\theta}/\left(  \omega_{\mathrm{a}}%
r_{0}\right)  $. With the help of Eq. (\ref{<xi_theta xi_theta>}) one finds
that%
\begin{equation}
\left\langle i_{\mathrm{n}}\left(  \tau\right)  i_{\mathrm{n}}\left(
\tau^{\prime}\right)  \right\rangle =\frac{2\Theta}{\omega_{\mathrm{a}}%
r_{0}^{2}}\delta\left(  \tau-\tau^{\prime}\right)  \;.
\label{<i_n(tau)*i_n(tau')>}%
\end{equation}

\subsection{The noiseless case}

Consider first the noiseless case, for which $i_{\mathrm{n}}=0$. Below the
time evolution of $\gamma$ in the region $\left\vert i_{\mathrm{b}}\right\vert
>1$ is derived. Rewriting Eq. (\ref{CBJJ V2}) as $\mathrm{d}\tau
=\mathrm{d}\gamma/\left(  i_{\mathrm{b}}-\sin\gamma\right)  $ leads by integration
 to%
\begin{equation}
\tau=\frac{2}{\sqrt{i_{\mathrm{b}}^{2}-1}}\tan^{-1}\frac{i_{\mathrm{b}}%
\tan\frac{\gamma}{2}-1}{\sqrt{i_{\mathrm{b}}^{2}-1}}\;. \label{tau(gamma)}%
\end{equation}
Inverting this relation yields%
\begin{equation}
\gamma=2\tan^{-1}\frac{1+\sqrt{i_{\mathrm{b}}^{2}-1}\tan\frac{\pi\tau
}{T_{\mathrm{J}}}}{i_{\mathrm{b}}}\;, \label{gamma(tau)}%
\end{equation}
where the normalized period time $T_{\mathrm{J}}$ is given by%
\begin{equation}
T_{\mathrm{J}}=\frac{2\pi}{\sqrt{i_{\mathrm{b}}^{2}-1}}\;, \label{CBJJ T}%
\end{equation}
and thus%
\begin{equation}
A_{\theta}=2\tan^{-1}\frac{1+\sqrt{i_{\mathrm{b}}^{2}-1}\tan\left(
\frac{\sqrt{i_{\mathrm{b}}^{2}-1}\omega_{\mathrm{a}}}{2\omega_{\mathrm{d}}%
}\omega_{\mathrm{d}}t\right)  }{i_{\mathrm{b}}}-\omega_{\mathrm{d}}t\;.
\end{equation}
The averaged derivative $\left\langle \mathrm{d}\gamma/\mathrm{d}%
\tau\right\rangle $\ is related to the period time $T_{\mathrm{J}}$ by%
\begin{equation}
\left\langle \frac{\mathrm{d}\gamma}{\mathrm{d}\tau}\right\rangle =\frac
{1}{T_{\mathrm{J}}}\int_{0}^{T_{\mathrm{J}}}\frac{\mathrm{d}\gamma}%
{\mathrm{d}\tau}\mathrm{d}\tau=\frac{2\pi}{T_{\mathrm{J}}}\;,
\end{equation}
thus [see Eqs. (\ref{i_b}) and (\ref{CBJJ T})]%
\begin{equation}
\left\langle \frac{\mathrm{d}\gamma}{\mathrm{d}\tau}\right\rangle
=\sqrt{i_{\mathrm{b}}^{2}-1}\;, \label{V_DC ODJJ}%
\end{equation}
and%
\begin{equation}
-\left\langle \frac{\mathrm{d}A_{\theta}}{\mathrm{d}t}\right\rangle =\left\{
\begin{array}
[c]{cc}%
\omega_{\mathrm{d}} & \left\vert i_{\mathrm{b}}\right\vert <1\\
\omega_{\mathrm{d}}-\omega_{\mathrm{a}}\sqrt{i_{\mathrm{b}}^{2}-1} &
\left\vert i_{\mathrm{b}}\right\vert \geq1
\end{array}
\right.  \;.
\end{equation}
Therefore, the signal $\cos A_{\theta}$ when $\left\vert i_{\mathrm{b}%
}\right\vert \geq1$\ has Fourier spectrum made of peaks at the angular
frequencies $\omega_{\mathrm{d}}+n\omega_{\mathrm{s}}$, where $n$ is an
integer and the sideband spacing $\omega_{\mathrm{s}}$, which is given by
$\omega_{\mathrm{s}}=2\pi\omega_{\mathrm{a}}/T_{\mathrm{J}}=\sqrt{\left(
\omega_{\mathrm{d}}-\Omega_{\mathrm{eff}}\right)  ^{2}-\omega_{\mathrm{a}}%
^{2}}$ [see Eqs. (\ref{i_b}) and (\ref{CBJJ T})], depends on both the
amplitude $\omega_{\mathrm{a}}$ and frequency $\omega_{\mathrm{d}}$ of the
forcing term $\xi=r_{0}\omega_{\mathrm{a}}A_{r}e^{-i\omega_{\mathrm{d}}t}$.

\subsection{Fourier expansion}

With the help of Eqs. (\ref{CBJJ V2}), (\ref{gamma(tau)}) and (\ref{CBJJ T})
together with the identity $\sin\left(  2\tan^{-1}\left(  s\right)  \right)
=2s/\left(  1+s^{2}\right)  $ one finds that%
\begin{equation}
\frac{\mathrm{d}\gamma}{\mathrm{d}\tau}=\mathcal{V}\left(  \frac{2\pi\tau
}{T_{\mathrm{J}}}+x_{0}\right)  \;, \label{d gamma/d tau}%
\end{equation}
where the function $\mathcal{V}\left(  x\right)  $ is defined by%
\begin{equation}
\mathcal{V}\left(  x\right)  =\frac{i_{\mathrm{b}}^{2}-1}{i_{\mathrm{b}}+\sin
x}\;, \label{V(x)=}%
\end{equation}
and where%
\begin{equation}
x_{0}=\tan^{-1}\frac{1}{\sqrt{i_{\mathrm{b}}^{2}-1}}\;. \label{x_0}%
\end{equation}
The Fourier expansion of the function $\mathcal{V}\left(  x\right)  $ is
expressed as [see Eq. (\ref{V(x)=})]%
\begin{equation}
\mathcal{V}\left(  x\right)  =\sum\limits_{k=-\infty}^{\infty}g_{k}e^{ikx}\;,
\label{V FE}%
\end{equation}%
where
\begin{equation}
g_{k}=\frac{i_{\mathrm{b}}}{\left\vert i_{\mathrm{b}}\right\vert }%
\sqrt{i_{\mathrm{b}}^{2}-1}i^{k}\left(  i_{\mathrm{b}}-\frac{i_{\mathrm{b}}%
}{\left\vert i_{\mathrm{b}}\right\vert }\sqrt{i_{\mathrm{b}}^{2}-1}\right)
^{\left\vert k\right\vert }\;. \label{g_k}%
\end{equation}
Thus, the magnitude of the sideband peaks is relatively large when $\left\vert
i_{\mathrm{b}}\right\vert \gtrsim1$. For $\left\vert i_{\mathrm{b}}\right\vert
\gg1$ the sidebands become small, and the relative phase becomes $\gamma
\simeq\left(  \omega_{\mathrm{d}}-\Omega_{\mathrm{eff}}\right)  t$ [see Eq.
(\ref{gamma(tau)})], i.e. the effect of laser modulation becomes weak.

\section{Square root law}

\label{App_SRL}

In each of the regions of phase locking \cite{Paciorek_1723,Adler_351} the
so-called winding number, which measures the oscillator phase accumulation relative to the phase of the drive, shows a plateau
\cite{Jensen_1637,Ben-Jacob_822,Reichhardt_414,Shim_95}. For concreteness, consider the region
just outside the primary plateau where (for positive $i_{\mathrm{b}}$)
$0<i_{\mathrm{b}}-1\ll1$, where $i_{\mathrm{b}}$ is the normalized detuning
(recall that $\left\vert i_{\mathrm{b}}\right\vert =1$ at the onset of
synchronization). The dimensionless frequency $T_{\mathrm{J}}^{-1}$ in this
region is related to the critical parameter $i_{\mathrm{b}}-1$ by
$T_{\mathrm{J}}^{-1}\simeq2^{-1/2}\pi^{-1}\sqrt{i_{\mathrm{b}}-1}$ [see Eq.
(\ref{CBJJ T})]. Below it is shown that a similar 'square root law' can be
obtained for the region just outside any other plateau.

Consider a general 1D map having the form%
\begin{equation}
\theta_{n+1}=\theta_{n}-\alpha+{W_{\mathrm{a}}\mathcal{F}\left(  \theta_{n}\right)
}\;,\label{Q_n+1 SRL}%
\end{equation}
that describes, e.g. the phase of the oscillator motion observed every period of the applied drive. The detuning $\alpha$ is approximately $(\Omega_{\mathrm{eff}}-\omega_{\mathrm{d}})2\pi/\omega_{\mathrm{d}}$. $W_{\mathrm{a}}$ parametrizes the strength of applied periodic drive (zero in the absence of drive). 
The ``map function" is a periodic real function, if we assume that the phase at the sampling step $n+1$ only depends on phase at step $n$, and the phases $\theta$ and $\theta + 2\pi$ are physically indistinguishable, $\mathcal{F}\left(  \theta+2\pi\right)
=\mathcal{F}\left(  \theta\right)  $. 

\begin{figure}
[ptb]
\includegraphics[
height=3.2019in,
width=3.3541in
]%
{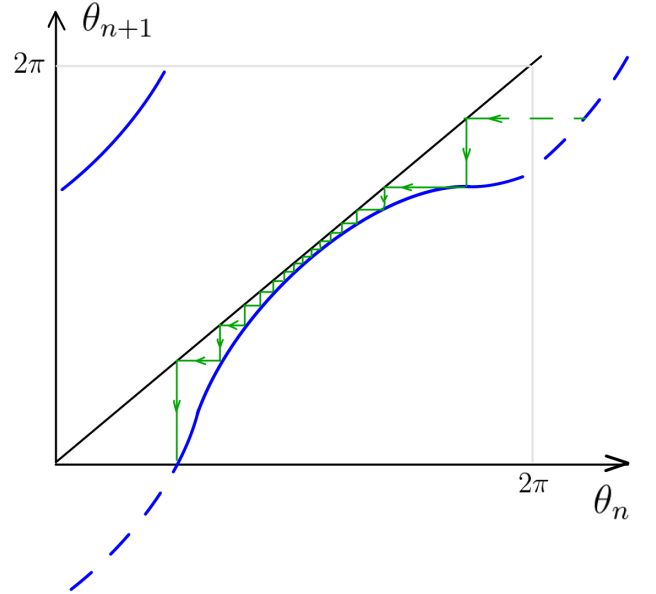}%

\caption{Phase locking-unlocking transition. Blue line represents mapping of Eq. (\ref{Q_n+1 SRL}), on the unlocked side of the transition. Green line is the staircase diagram that bounces between the map function and the diagonal $\theta_{n+1} = \theta_n$, graphically implementing the iteration procedure. Note that most of the iterations (longest time) is spend around the near-touching region. This leads to the ``square-root" law for the splitting between the drive frequency and the oscillator frequency.}
\end{figure}

If the drive frequency is near the oscillator frequency, then the detuning is close to zero.  The phase locking corresponds to a stable fixed point of the map, which can be found by solving the equation $\theta_{n+1} =\theta_{n}\equiv \theta^*$, that is 
\begin{equation}
\alpha={W_{\mathrm{a}}\mathcal{F}\left(  \theta^*\right)
}\;,\label{Q_n+1 SRLf}%
\end{equation}
where $\theta^*$ is the stroboscopic value of  the oscillator phase at the fixed point.
Since the function $\cal F$ is continuous and periodic, it is bounded, and the fixed point of this map only exists for a small enough detuning.

Suppose now that the drive frequency is an integer fraction of the oscillator frequency, $\omega_{\mathrm{d}} \approx  \Omega_{\mathrm{eff}}/p$. Then, since the phase is defined modulo $2\pi$, the effective detuning in this case is still small and the analysis of the locking transition is based on the same map equation. The locking will occur to $\Omega_{\mathrm{eff}} = p\omega_{\mathrm{d}}$ On the other hand, if the frequency of the drive is an integer multiple ($\omega_{\mathrm{d}} \approx p \Omega_{\mathrm{eff}}$) of the oscillator frequency, then during one period of the drive, oscillator will only accrue a fraction  $\approx 2\pi/p$ phase, which in general will correspond to too large of a detuning over single period of the drive to allow for a fixed point solution. 
However, the problem can be reduced to a formally equivalent one by iterating the map $p$ times,

\begin{equation}
\theta_{n+p}=\theta_{n}-p\alpha+{\tilde W_{\mathrm{a}}\mathcal{\tilde F}\left(  \theta_{n}\right)
}\;,\label{Q_n+p SRL}%
\end{equation}
where the effective detuning $p\alpha$ is now close to $2\pi$, and the original treatment for the simple locking applies, with the resulting locked frequency being $\Omega_{\mathrm{eff}} = \omega_{\mathrm{d}}/p$. Note however, that the amplitude and the form of the map function are now different, and generically smaller in magnitude that the $p = 1$ case. This makes locking to high frequency drives in general less stable.

Now, let us look more carefully at the transition out of the locking regime as a function of detuning, $\alpha$. Let us define $\alpha_c$ as the  value of detuning and $\theta_c$ as the value of the fixed point $\theta^*$ at the transition between locked and the running phase (unlocked) regimes, $\theta_c = \theta^*(\alpha_c)$. At the critical detuning, the slope of the map function $\cal F$ vanishes (see Figure 4).  We can expand the map function  around this point up to the second order, 
\begin{equation}
\mathcal{F}\left(  \theta\right)  =\frac{\alpha_{\mathrm{c}}%
}{W_{\mathrm{a}}}\left[  1-z^2\left(  \theta-\theta_{\mathrm{c}}\right)  ^{2}\right]
+O\left(  \left(  \theta-\theta_{\mathrm{c}}\right)  ^{3}\right)  \;,\label{F 2nd order}%
\end{equation}
where $z$ is a parameter that characterizes the curvature of the map function near $\theta_c$.
 Indeed, as can be seen from Eq.
(\ref{Q_n+1 SRL}), the map has a fixed point (i.e. a solution to the equation
$\theta_{n+1}=\theta_{n}$) provided that $\alpha<\alpha_{\mathrm{c}}$ (the fixed point
is given by $\theta^*=\theta_{\mathrm{c}}+\sqrt{1-\alpha/\alpha_{\mathrm{c}}}/z$).

On the unlocked side, it is convenient to treat the discreet series $y_{n}=\theta_{n}-\theta_{\mathrm{c}}$ as a continuous
function $y\left(  n\right)$, since near $\theta_c$ each iteration step changes $\theta$ only slightly (``staircase" diagram in Figure 4). Then, from 
(\ref{Q_n+1 SRL}) and (\ref{F 2nd order})%
\begin{equation}
\frac{\mathrm{d}y}{\mathrm{d}n}=\alpha_{\mathrm{c}}-\alpha-\alpha_{\mathrm{c}%
}z^2y^{2}\;.
\end{equation}
This equation can be easily integrated in order to determine how long it takes the phase to pass through the bottleneck (Fig. 4). For that, the limits of integration over $y$ can be extended to the full real axis (details of the map function away from $\theta_c$ do not matter in this regime, since it takes much shorter time to go through regions other than the bottleneck).

Thus the estimate for the total number of drive periods for the phase to wind by $2\pi$ relative to the drive, is

\begin{align}
N(\alpha) &  =\frac{1}{\alpha-\alpha_{\mathrm{c}}}\int_{-\infty}^{\infty}%
\frac{\mathrm{d}y}{1+\frac{\alpha_{\mathrm{c}}}{\alpha-\alpha_{\mathrm{c}}%
}z^2y^{2}}\nonumber\\
&  =\frac{\pi  }{z\sqrt{\alpha_{\mathrm{c}}
|\alpha-\alpha_{\mathrm{c}}|  }}\;.\nonumber
\end{align}
This translates into the time period $T_{\mathrm{LC}}=2\pi N(\alpha)/\omega_d$ per $2\pi$ phase slip. 
Thus, the periodic oscillation of $y$ gives rise to sideband spectral peaks split from the drive frequency by integer multiples of  $2\pi/T_{\mathrm{LC}}\propto \sqrt{|\alpha
-\alpha_{\mathrm{c}}|}$.

\newpage
\bibliographystyle{IEEEtran}
\bibliography{Eyal_Bib}

\end{document}